 \documentclass[final,5p,times,twocolumn]{elsarticle}

\usepackage{graphicx}
\usepackage{amssymb}
\usepackage{amsmath}

\biboptions{comma,square}

\journal{Physica E}

\begin{document}

\begin{frontmatter}



\title{Crossover of the Hall-voltage distribution in AC quantum Hall effect}


\author{H. Akera}

\address{Division of Applied Physics, Faculty of Engineering, 
Hokkaido University, Sapporo, Hokkaido, 060-8628, Japan}

\begin{abstract}
The distribution of the Hall voltage induced by low-frequency AC current
is studied theoretically 
in the incoherent linear transport of quantum Hall systems. 
It is shown that the Hall-voltage distribution makes a crossover from 
the uniform distribution to a concentrated-near-edges distribution 
as the frequency is increased or the diagonal conductivity is decreased. 
This crossover is also reflected in 
the frequency dependence of AC magnetoresistance. 
\end{abstract}


\end{frontmatter}

\def\sxx{\sigma_{xx}}
\def\sxy{\sigma_{xy}}
\def\syx{\sigma_{yx}}
\def\syy{\sigma_{yy}}
\def\sab{\sigma_{\alpha \beta}}
\def\cab{\chi_{\alpha \beta}}
\def\cyy{\chi_{yy}}
\def\cyx{\chi_{yx}}
\def\oc{\omega_{\rm c}}
\def\ve{\varepsilon}
\def\muec{\mu_{\rm ec}}
\def\kB{k_{\rm B}}
\def\Vec#1{\mbox{\boldmath $#1$}}

\section{Introduction}

In the quantum Hall effect \cite{Klitzing1980,Kawaji1981} observed 
in two-dimesional electron systems (2DES) in strong magnetic fields, 
the Hall voltage $V_H$ divided by the current $I$ is quantized as 
\begin{equation}
\frac{V_H}{I} = \frac{h}{ie^2}
\label{eq:QHE}
\end{equation}
with $i$ an integer. 
The distribution of this quantized Hall voltage along the width of the 2DES 
has been studied theoretically and experimentally, 
but has a problem that remains to be solved. 

MacDonald, Rice and Brinkman \cite{MacDonald1983} have studied theoretically 
the Hall-voltage distribution in the ideal 2DES with no disorder 
for integer values of the Landau-level filling factor $\nu$ at absolute zero. 
They have considered an infinitely-long sample with width $W$ 
in the $xy$ plane 
in the magnetic field $B$ along the $z$ direction ($B>0$). 
The 2DES considered has a macroscopic size:   
$W$ is much larger than the magnetic length $l=(\hbar c/eB)^{1/2}$ ($e>0$).  
In this paper we choose the $x$ axis along the current 
and the $y$ axis along the width (the 2DES is in $-W/2 <y< W/2$).
In the ideal 2DES with a constant current, 
the electric field along the current $E_x$ is zero 
and the dissipation is absent. 
The Hall field $E_y(y)$ induces a shift of each wave function by 
$\Delta y = - e E_y/m \oc^2$ with $\oc = eB/mc$ ($m$: the effective mass),  
and the resulting polarization gives 
the Hall charge density $\rho_{\rm polar}(y)$. 
MacDonald et al. \cite{MacDonald1983} have obtained 
the formula for $\rho_{\rm polar}(y)$ 
by making the summation of 
contributions from each of these shifted wave functions. 
The same formula is obtained by starting with   
the polarization or 
the dipole moment per area which is given by 
\begin{equation}
P_y = -e \Delta y \nu/2\pi l^2 = \cyy^0 E_y ,
\label{eq:polarization}
\end{equation}
where $\cyy^0$ is the DC dielectric susceptibility 
(the superscript 0 means DC) given by 
\begin{equation}
\cyy^0=e^2 \nu/ h\oc .
\label{eq:susceptibility}
\end{equation}
With use of $\rho_{\rm polar}= - \nabla_y P_y$ ($\nabla_y=\partial/\partial y$), 
we obtain 
\begin{equation}
\rho_{\rm polar} 
= - \cyy^0 \nabla_y E_y
=  \cyy^0 \nabla_y^2 \phi . 
\label{eq:rho_polar}
\end{equation}
The electrostatic potential $\phi(y)$ ($E_y=-\nabla_y \phi$)  
in this equation is given in terms of $\rho_{\rm polar}(y)$ by
\begin{equation}
\phi(y)= -\frac{2}{\ve} \int_{-W/2}^{W/2} dy' \ln |y-y'| \rho_{\rm polar}(y') 
\label{eq:long_range_pot}
\end{equation}
where $\ve$ is the dielectric constant of the intrinsic semiconductor. 
Equations (\ref{eq:rho_polar}) and (\ref{eq:long_range_pot}) 
give the Hall potential $\phi(y)$ as a function of $y$.  
The calculated result \cite{MacDonald1983} shows that 
\textit{the Hall voltage is concentrated near edges}.  

In the presence of dissipation the Hall-voltage distribution 
changes drastically \cite{Thouless1985}. 
Here we assume that the transport current densities $j_x$ and $j_y$ 
are related to $E_x$ and $E_y$ 
by the local DC conductivity tensor $\sxx^0=\syy^0$, $\sxy^0=- \syx^0$, 
which has no spatial dependence, that is, 
\begin{equation}
j_x=\sxx^0 E_x + \sxy^0 E_y , \ \ 
j_y=\syx^0 E_x + \syy^0 E_y .
\label{eq:jx_jy}
\end{equation}
The density of the charge accumulated due to the transport, 
$\rho_{\rm trans}$, evolves according to the equation of charge conservation:  
\begin{equation}
\frac{\partial \rho_{\rm trans}}{\partial t} 
= - \Vec \nabla \cdot \Vec j
= - \syy^0 \Vec{\nabla} \cdot \Vec{E} .
\end{equation}
If we consider a state which is steady and uniform along $x$, 
the above equation shows that 
\textit{the Hall field is uniform along the width}. 
The uniform Hall field means a uniform current density, which is along the $x$ direction.  
Such distributions of the Hall voltage and the current are 
those minimizing the total entropy production, 
which is in accordance with 
the theorem of the minimum entropy production \cite{GrootMazur1962}.

Many other theoretical works 
have been performed 
on the Hall-voltage and current distributions 
both in the absence and in the presence of dissipation.  
In the dissipationless case, 
quantum wires with width comparable to $l$ 
have been studied by calculating the wave function numerically 
and taking into account $\rho_{\rm polar}$ 
in this way \cite{Heinonen1985,Pfannkuche1992,Wexler1994}. 
In several papers \cite{Thouless1993,Wexler1994,Hirai1994}  
the edge charge due to electrons added to 
(and subtracted from) edge states was considered. 
Such edge charge can be described as 
the charge due to the polarization in eq.(\ref{eq:rho_polar})
since adding and subtracting electrons in this way 
is equivalent to shifting the whole electrons by the appropriate distance. 
The theory has also been extended 
to the fractional quantum Hall states \cite{Palacios1998}. 
In the dissipative case, 
the theory has been extended to 
a state with compressible and incompressible strips 
in a slowly-varying confining potential \cite{Guven2003,Siddiki2004,Kanamaru2006}. 

Fontein et al. \cite{Fontein1991,Fontein1992} have measured 
the Hall potential in a 2-mm-wide 2DES 
formed in a GaAs/AlGaAs heterostructure
using the linear electro-optic effect. 
They have observed a crossover of the Hall-voltage distribution 
from the \textit{concentrated-near-edges} to the \textit{uniform} distribution 
by increasing the temperature or the current.  
This observation suggests that the crossover occurs 
with increasing the dissipation $\sxx^0$. 
At first glance this seems to contradict the expectation from the above theories: 
any real systems of macroscopic size should have nonzero dissipation 
and should show the uniform distribution.
A possible reason for the contradiction 
may be the difference in angular frequency $\omega$ of current:   
the theories assumed $\omega=0$ (the steady state), 
while the experiment applied AC current of $\omega/2\pi = 235$Hz 
to employ the lock-in technique.  
The theorem of the minimum entropy production \cite{GrootMazur1962}, 
which leads to the uniform distribution, 
is applicable only to the case of $\omega = 0$. 

In this paper we study theoretically the Hall-voltage distribution 
in the case of $\omega \not= 0$. 
The 2DES we consider here is uniform except at sharp edges,  
while a 2DES with a slowly-varying confining potential 
will be studied elsewhere. 
The value of the filling factor in the uniform bulk region 
is not restricted to integers,  
but we neglect the electron correlation 
such as in the fractional quantum Hall effect 
by considering a relatively high temperature. 
In this paper we study only the incoherent linear transport   
by employing the local conductivity tensor. 
A crossover between coherent and incoherent regimes 
has been studied theoretically \cite{Ando1996,Ando1998} 
for the voltage distribution in the 2DES with source and drain contacts 
in strong magnetic fields. 

The organization of the paper is as follows. 
In \S 2, we introduce a model and 
derive an equation for the Hall potential as a function of $y$. 
In \S 3, we present an analytical solution for the Hall potential 
when complex conductivities are constant and the interaction is short-ranged. 
In \S 4, we study numerically the Hall potential in the long-range interaction 
as well as in the short-range interaction.  
We present 
our model for the complex conductivities in the edge region, 
our method of numerical calculation, and calculated results.  
In \S 5, conclusions and discussion are given. 
In Appendix we estimate the value of the complex conductivities. 

\section{Model and Equations}
\subsection{Current Density and Complex Conductivity}

We assume that the response of the current to the electric field is local. 
That is, 
the current density $j_{\alpha}(\Vec r, t)=j_{\alpha}(\Vec r, \omega)e^{i \omega t}$ 
at the position $\Vec r=(x,y)$ 
is determined only by 
the electric field at the same position $E_{\beta}(\Vec r, \omega)e^{i \omega t}$ 
($\alpha,\beta=x,y$):  
\begin{equation}
j_{\alpha}(\Vec r, \omega)= 
\sum_{\beta} \sab(\Vec r, \omega) E_{\beta}(\Vec r, \omega) .
\label{eq:ACj}
\end{equation}
This local relation may be applicable to macroscopic samples 
where $W \gg l_{\phi}$ 
with $l_{\phi}$ the phase coherence length,  
because in this case 
the length scale of variations of the electric field,  
which is of the order of $W$, is much larger than $l_{\phi}$.  
The conductivity $\sab(\Vec r, \omega)$ in this case can be determined 
by calculating the uniform-current response to the uniform electric field 
and taking the average over the random potential 
with the length scale $l_{\rm ran}$ 
since we assume here that $l_{\rm ran}< l_{\phi}$. 
This averaging procedure makes the 2DES isotropic in the $xy$ plane
so that we have 
$\sxx(\Vec r, \omega)=\syy(\Vec r, \omega)$ and 
$\sxy(\Vec r, \omega)=-\syx(\Vec r, \omega)$. 
The spatial dependence of $\sab(\Vec r, \omega)$ in this paper is due to 
the decrease of the local electron density 
as approaching a boundary of the 2DES.  
When $\omega \not= 0$, 
the above relation eq.(\ref{eq:ACj}) 
can be rewritten, in terms of 
the polarization $P_{\alpha}(\Vec r, \omega)=j_{\alpha}(\Vec r, \omega)/i\omega$  
and the dielectric susceptibility 
$\cab(\Vec r, \omega)=\sab(\Vec r, \omega)/i\omega$, as  
$P_{\alpha}(\Vec r, \omega)= 
\sum_{\beta} \cab(\Vec r, \omega) E_{\beta}(\Vec r, \omega)$. 

Now we restrict our discussion to the low-frequency region.  
The relevant energy scales in the response to the AC electric field are 
$\hbar \oc$ and the Landau-level broadening due to the random potential. 
By assuming that $\hbar \omega$ is much smaller than such energy scales, 
we expand $\sab(\omega)$ in a power series of $\omega$ 
and retain terms up to the first order of $\omega$. 
Since the real and imaginary parts of 
$\sab(\omega)=\sab'(\omega) +i\sab''(\omega)$ satisfy the following relation: 
$\sab'(-\omega) = \sab'(\omega)$ and $\sab''(-\omega) = - \sab''(\omega)$, 
we can write $\sab(\omega)$ as
\begin{equation}
\sab(\omega) = \sab^0 + i \omega \cab^0 ,
\label{eq:complex_sigma}
\end{equation}
where $\sab^0$ is the DC conductivity 
and $\cab^0$ is the DC susceptibility.  

\subsection{Hall Charge Density}

In this paper we consider 
a 2DES with two boundaries which are both parallel to the $x$ axis. 
We assume that $\sab$ and $E_{\beta}$ are uniform along $x$. 
In the vicinity of the boundaries, $\sab$ has a $y$ dependence,  
which will be specified in \S \ref{sec:Model_for_ydep}.
From the equation of charge conservation, 
the Hall charge density $\rho$ is given by  
\begin{equation}
i \omega \rho(y) = - \nabla_y j_y (y) .
\label{eq:charge_conservation}
\end{equation}
In this equation and in the following, 
the $\omega$ dependence of the variables and coefficients 
will not be shown explicitly.
From eq.(\ref{eq:ACj}), we have
\begin{equation}
j_y (y) = \syx(y) E_x + \syy(y) E_y(y) .
\label{eq:jy}
\end{equation}
Here we have also assumed that $E_x$ has no dependence on $y$ 
since $\nabla_y E_x -\nabla_x E_y \approx 0$ when $\omega$ is small. 
The above equations show that, 
when $E_x \not= 0$, the $y$ dependence of $\syx$ gives 
the Hall charge density $\rho$ and the Hall field $E_y$.

\subsection{Current due to the Chemical-Potential Gradient}

Corresponding to the two terms of $\sab$ in eq.(\ref{eq:complex_sigma}), 
$\rho(y)$ has two components:  
$\rho(y) = \rho_{\rm trans}(y)  + \rho_{\rm polar}(y)$,
where $\rho_{\rm trans}$ is the transport charge density defined by
\begin{equation}
i \omega \rho_{\rm trans}(y) = - \nabla_y  
\left[ \syx^0(y) E_x 
+ \syy^0(y) E_y(y) \right] .
\label{eq:rho_trans}
\end{equation}
and $\rho_{\rm polar}$ is the polarization charge density defined by 
\begin{equation}
\rho_{\rm polar}(y) = - \nabla_y  
\left[ \cyx^0(y) E_x 
+ \cyy^0(y) E_y(y) \right] .
\end{equation}

The transport charge density $\rho_{\rm trans}(y)$ 
gives a deviation of the chemical potential $\mu$ 
from its equilibrium value $\mu_{\rm eq}$. 
The deviation $\Delta \mu= \mu - \mu_{\rm eq}$ is given by 
\begin{equation}
\Delta \mu(y) = \rho_{\rm trans}(y)/(-eD_{\rm T}) ,
\label{eq:Delta_mu}
\end{equation}
where $D_{\rm T}=\partial n/\partial \mu$ 
($n$: electron density per unit area) is the thermodynamic density of states. 
The gradient of $\Delta \mu$ induces the current and 
the total current density is given in terms of 
the gradient of the electrochemical potential ($\muec$), 
$\nabla_y \muec=e E_y + \nabla_y \Delta \mu$, as
\begin{equation}
j_y (y) = \syx (y) E_x 
+ \syy^0(y) e^{-1} \nabla_y \muec + i\omega \cyy^0(y) E_y(y) .
\end{equation}
Note that $\nabla_x \Delta \mu =0$ and  
the polarization current is induced only by the electric field. 
We have calculated numerically the value of $\nabla_y \Delta \mu$ 
and have obtained $|\nabla_y \Delta \mu| \ll e |E_y|$,  
which is also supported by an analytical result below in eq.(\ref{eq:nabla_mu}).
Therefore we will neglect the term proportional to $\nabla_y \Delta \mu$ 
in the following. 

\subsection{Hall Potential}

The electrostatic potential due to the Hall charge, $\phi(y)$, 
in a 2DES uniform along $x$ is given by 
\begin{equation}
\phi(y) = \int_{-\infty}^{\infty} dy' K(y-y') \rho(y') . 
\label{eq:electrostatic}
\end{equation}
where $K(y-y')$ is the potential due to the unit line charge at a distance $|y-y'|$. 
We consider the two models of the electrostatic interaction.  
One is the long-range interaction with  
\begin{equation}
K(y-y') = - \frac{2}{\ve} \ln |y-y'| . 
\label{eq:K_long_range}
\end{equation}
This is the potential in a dielectric material 
with the dielectric constant $\ve$ 
and has been used in the previous work by MacDonald et al. \cite{MacDonald1983}. 

The other model is the short-range interaction with  
\begin{equation}
K(y-y') = r_K \delta(y-y') . 
\label{eq:K_short_range}
\end{equation}
In this model 
$\phi(y) = r_K \rho(y)$.
This model is valid when 
the range of $K(y-y')$ is much shorter than 
the length scale of variation of $\rho(y)$, $L_{\rho}$.  
If we consider a 2DES with a parallel gate electrode at distance $d$, 
this condition becomes $d \ll L_{\rho}$. 
In such system 
\begin{equation}
r_K=4\pi d/\ve . 
\label{eq:rK}
\end{equation}

\section{Short-Range Interaction and Constant Conductivity}
\label{sec:Short-Range}

We first consider the simpler case of the short-range interaction.  
If the interaction is short-ranged, 
electrostatics, in addition to transport, becomes local 
and the Hall potential $\phi(y)$ is described by a differential equation. 
In this section 
we consider the bulk uniform region, as the simplest case,  
where complex conductivities $\syx$ and $\syy$ have no spatial dependence 
and $\syx = \syx^{0\rm b} + i\omega \cyx^{0\rm b}$,  
$\syy = \syy^{0\rm b} + i\omega \cyy^{0\rm b}$. 
In this case $\phi(y)$ is 
described by a differential equation with constant coefficients:  
\begin{equation}
i \omega \phi(y) 
= D (1+ i\tilde \omega) \nabla_y^2 \phi(y) ,
\label{eq:c_diffusion_eq}
\end{equation}
where $D$ is a diffusion constant given by
\begin{equation}
D= r_K \syy^{0\rm b} , 
\label{eq:diffusion_const}
\end{equation}
and $\tilde \omega$ is a normalized angular frequency defined by
\begin{equation}
\tilde \omega =  \omega \cyy^{0\rm b} / \syy^{0\rm b}  .
\label{eq:omega_tilde}
\end{equation}

When $\tilde \omega \ll 1$, the equation for $\phi(y)$ becomes
\begin{equation}
i \omega \phi(y) 
= D  \nabla_y^2 \phi(y) .
\end{equation}
Then $\phi(y)$ is given by
\begin{equation}
\phi(y) = \phi_0 \exp[-(1+i)y/\lambda(\tilde \omega) ] , 
\end{equation}
and is decaying and oscillating with $y$. 
We also have a solution: $\phi(y) = \phi_0\exp[(1+i)y/\lambda ]$. 
Here $\lambda(\tilde \omega)$ is the decay length given by
\begin{equation}
\lambda(\tilde \omega) = \sqrt{2D/\omega} ,
\label{eq:decay_length1}
\end{equation}
and is equal to the diffusion length in the time interval $1/\omega$. 
In this section we consider the change of $\lambda$ 
when either $\omega$ or $\syy^{0\rm b}$ is changed. 
Since $\lambda(\tilde \omega) = \sqrt{2 r_K \cyy^{0\rm b}/\tilde \omega}$, 
$\lambda(\tilde \omega)$ decreases with the increase of $\tilde \omega$, 
that is $\lambda(\tilde \omega)$ decreases when $\omega$ is increased or 
$\syy^{0\rm b}$ is decreased. 

When $\tilde \omega \gg 1$, the equation for $\phi(y)$ becomes
\begin{equation}
\phi(y) 
= r_K \cyy^{0\rm b} \  \nabla_y^2 \phi(y) .
\end{equation}
The decaying solution in this case is
\begin{equation}
\phi(y) = \phi_0 \exp(- y/\lambda_{\infty} ) ,
\end{equation}
and the decay length is
\begin{equation}
\lambda_{\infty} = \sqrt{r_K \cyy^{0\rm b} }  .
\label{eq:decay_length2}
\end{equation}
If we employ eq.(\ref{eq:rK}) and an estimate using eq.(\ref{eq:estimate}), 
$\cyy^{0\rm b}=e^2 \nu_{\rm b}/(h\oc)$ 
with $\nu_{\rm b}$ the bulk filling factor, 
we have $\lambda_{\infty} = \sqrt{2\nu_{\rm b} l^2 d /a_{\rm B}^*}$  
where $a_{\rm B}^* = \hbar^2 \ve /me^2$ is the effective Bohr radius. 
If we use the value of $m$ and $\ve$ of GaAs, $\nu_{\rm b}=4$, $d=0.1\mu$m 
and $B=5$T, 
we obtain $\lambda_{\infty} \sim d$  
which means that the spatial variation of $\phi$ and $\rho$ in this case 
is too steep to satisfy 
the condition ($\lambda_{\infty} \gg d$) for the short-range model. 
In the lower-magnetic-field region, however,  
$\lambda_{\infty}$ ($\propto B^{-1}$) becomes larger and
the condition becomes satisfied. 

Starting from $\tilde \omega =0$, we increase $\tilde \omega$.  
Then we first encounter a crossover 
around the point satisfying $\lambda(\tilde \omega) = W$ 
when $W$ is large enough. 
In this crossover 
the Hall-voltage distribution changes from \textit{uniform}
to \textit{concentrated-near-edges} profile. 
If we increase $\tilde \omega$ further, 
we come to another crossover  
around the point satisfying $\tilde \omega = 1$, 
where 
the dominant response changes from \textit{transport current} 
to \textit{polarization current}.
In the second crossover ($\tilde \omega = 1$)
the decay length becomes of the order of $\lambda_{\infty}$.  
To distinguish the second crossover from the first one, 
$\lambda_{\infty} \ll W$ must be satisfied,  
in addition to the condition 
for the short-range model of $\lambda_{\infty} \gg d$. 


We examine the validity of the approximation to neglect 
the current due to the chemical-potential gradient 
in the present uniform case in the short-range model
by showing $|\nabla_y \Delta \mu| \ll e |E_y|$. 
Using eqs.(\ref{eq:Delta_mu}), (\ref{eq:rho_trans}), 
(\ref{eq:c_diffusion_eq}) and (\ref{eq:diffusion_const}), 
we have
\begin{equation}
\frac{|\nabla_y \Delta \mu| }{ |e E_y|} 
= \left(e^2 D_{\rm T} \ r_K \sqrt{1+\tilde \omega^2} \right)^{-1}
\sim \frac{l}{d} ,
\label{eq:nabla_mu}
\end{equation}
where we have used $D_{\rm T} \sim 1/(2 \pi l^2 \hbar \oc)$ as well as 
eq.(\ref{eq:rK}) and $\tilde \omega \sim 1$. 
For $d/l=10$, $|\nabla_y \Delta \mu| / |e E_y| \sim 1/10$ 
and we can neglect $\nabla_y \Delta \mu$. 

\section{Numerical Calculation}
\subsection{Model for $\syx$ and $\syy$ in the edge region}
\label{sec:Model_for_ydep}

In the long-range interaction, 
the relation between the Hall potential $\phi(y)$ and 
the Hall charge density $\rho(y)$ is nonlocal. 
Therefore, the value of $\phi(y)$ in the bulk region is 
influenced by that of $\rho(y)$ in the edge region, 
which is in turn determined by 
eqs.(\ref{eq:charge_conservation})(\ref{eq:jy}) 
with $\syx(y)$ and $\syy(y)$ in the edge region. 
Here we introduce a model of 
$\syx(y)=\syx^0(y)+i\omega \cyx^0(y)$ and $\syy(y)=\syy^0(y)+i\omega \cyy^0(y)$ 
in the edge region.

In the edge region the electron density  
and the equilibrium chemical potential $\mu_{\rm eq}(y)$ decrease 
as approaching the boundary from the bulk region. 
We describe this $y$ dependence by a simple function: 
\begin{equation}
\begin{split}
\mu_{\rm eq}(y)&= \mu_{\rm b} , \ \ (0<y<W/2) \\
\mu_{\rm eq}(y)&= \mu_{\rm b} \left[1-(y-W/2)^2/W_{\rm e}^2 \right], \ \ (y>W/2)
\end{split}
\end{equation}
and $\mu_{\rm eq}(-y)=\mu_{\rm eq}(y)$. 
The parameter $W_{\rm e}$ represents the width of the edge region 
since $\mu_{\rm eq}(W/2+W_{\rm e})=0$.

We assume that 
the $y$ dependence of $\syx(y)$ and $\syy(y)$ originates from 
the $y$ dependence of $\mu_{\rm eq}(y)$, 
that is $\syx(y)=\syx(\mu_{\rm eq}(y))$ and $\syy(y)=\syy(\mu_{\rm eq}(y))$. 
As for the $\mu_{\rm eq}$ dependence of $\syx^0$ and $\syy^0$,  
we employ a model \cite{Akera2005} which retains 
the observed features of $\syx^0(B)$ and $\syy^0(B)$:
\begin{equation}
\syx^0(\mu_{\rm eq})= \frac{2e^2}{h} \sum_N f_N, 
\end{equation}
\begin{equation}
\syy^0(\mu_{\rm eq})  = 
\frac{2e^2 D_0}{\kB T} \sum_N  (2N+1) f_N \left( 1-f_N \right), 
\end{equation}
\begin{equation}
f_N=
\left\{ 1+
\exp\left[\left(\ve_N -\mu_{\rm eq} \right)/\kB T \right]
\right\}^{-1}
\end{equation}
where $\ve_N=\hbar \oc (N+1/2)$, $N=0,1,2,\cdots$, 
$T$ is the temperature and 
$D_0$ is a constant. 
As for $\cyx^0(\mu_{\rm eq})$ and $\cyy^0(\mu_{\rm eq})$ 
we use a simple formula 
\begin{equation}
\cyx^0(\mu_{\rm eq})=0, \ \ \ \cyy^0(\mu_{\rm eq})= e^2 \nu / (h \oc) ,
\label{eq:cyy}
\end{equation}
which is derived in Appendix. 
The derivation assumes that Landau-level mixings are negligible 
and also that $\Gamma$ is negligible compared to $\hbar \oc$ 
with $\Gamma$ the Landau-level broadening. 
Since such assumptions are not always satisfied in the cases we consider below,  
the above formula itself should be considered an assumption. 
Note that the conclusion of this paper does not change 
even when $\syx^0(\mu_{\rm eq})$, $\syy^0(\mu_{\rm eq})$ and $\cyy^0(\mu_{\rm eq})$  
change substantially, as will be shown below. 
In calculating the filling factor $\nu$ at a given $\mu_{\rm eq}$ and $T$, 
we use the following density of states: 
\begin{equation}
\begin{split}
D(\ve)&= 1/(2 \pi l^2 \Gamma) , \ \ \ \ \ (|\ve-\ve_N|<\Gamma) \\
D(\ve)&= 0, \ \ \ \ \ ({\rm otherwise})
\end{split}
\end{equation}

\begin{figure}[ht]
  \begin{center}
  \includegraphics[height=12cm, bb=0 0 595 840]{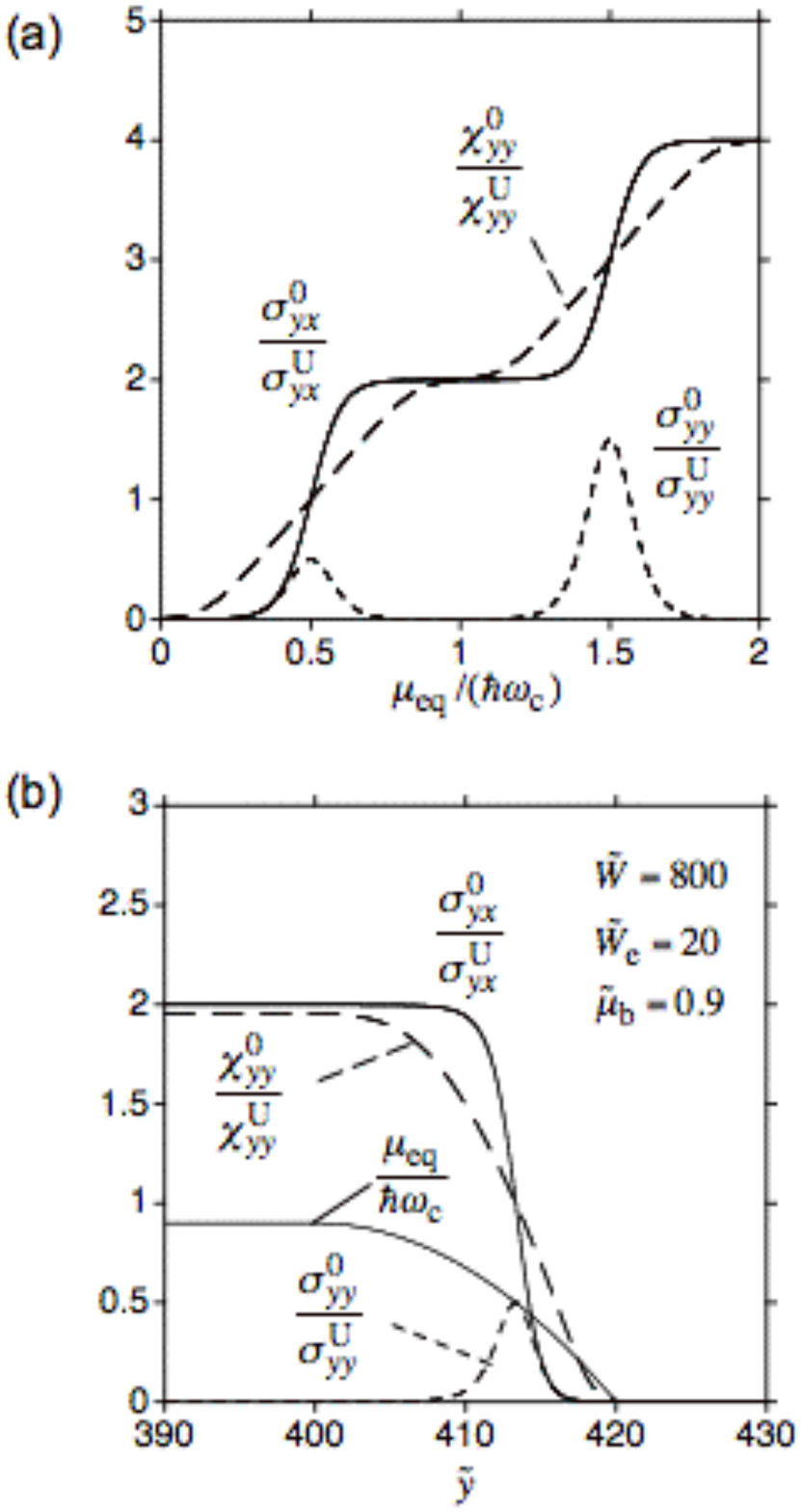}
  \end{center}
  \vskip -0.8cm
  \caption{
  (a) Dependence of $\syx^0$, $\syy^0$ and $\cyy^0$ 
  on the equilibrium chemical potential, $\mu_{\rm eq}$. 
  $\syx^{\rm U}=e^2/h$, $\syy^{\rm U}=e^2 D_0/(\kB T)$ and 
  $\cyy^{\rm U}=e^2/(h\oc)$. 
  (b) $y$ dependence of $\syx^0$, $\syy^0$ and $\cyy^0$ 
  as well as $\mu_{\rm eq}$. 
  $\tilde y=y/l_{\rm U}$, $\tilde W=W/l_{\rm U}$, $\tilde W_{\rm e}=W_{\rm e}/l_{\rm U}$ 
  and $\tilde \mu_{\rm b}=\mu_{\rm b}/(\hbar\oc)$ 
  with $l_{\rm U}$ defined in eq.(\ref{eq:lp}).
  }
  \label{fig:syy}
\end{figure}

Figure \ref{fig:syy}(a) presents 
the $\mu_{\rm eq}$ dependences of $\syx^0$, $\syy^0$ and $\cyy^0$, while  
Fig.\ref{fig:syy}(b) shows 
an example of the $y$ dependences of $\syx^0$, $\syy^0$ and $\cyy^0$ 
as well as that of $\mu_{\rm eq}$. 
In the numerical calculation 
the values of $\Gamma$ and $T$ are fixed as 
$\Gamma = 0.35 \hbar\oc$ and $\kB T = 0.05 \hbar\oc$. 


\subsection{Method of Numerical Calculation}

We calculate the Hall potential $\phi(y)$ and the Hall charge density $\rho(y)$
by solving eqs.(\ref{eq:charge_conservation}) and (\ref{eq:electrostatic}) 
with eq.(\ref{eq:K_long_range}). 
We consider a periodic array of infinitely-long 2DES strips. 
The $n$th strip is in $-W/2 +n W_{\rm p} <y< W/2 +n W_{\rm p}$ 
where $n$ is the integer and $W_{\rm p}$ is the periodicity.   
The Hall potential satisfies
\begin{equation}
\phi(-y)= -\phi(y), \ \ 
\phi(y+W_{\rm p})= \phi(y), 
\end{equation}
which leads to 
$\phi(y)= 0$ at $y=\pm W_{\rm p}/2$. 
The same is the case for $\rho(y)$. 
Therefore we expand $\phi(y)$ and $\rho(y)$ in the Fourier series as 
\begin{equation}
\phi(y)= \sum_{k=1}^{k_{\rm max}} \phi_k 
\sin \left( \frac{2\pi k}{W_{\rm p}} y \right) , \ \ \ 
\rho(y)= \sum_{k=1}^{k_{\rm max}} \rho_k 
\sin \left( \frac{2\pi k}{W_{\rm p}} y \right) .
\end{equation}
Then eqs.(\ref{eq:charge_conservation}) and (\ref{eq:electrostatic}) 
become a system of linear equations for $\phi_k$ and $\rho_k$ 
with a nonhomogeneous term 
proportional to $E_x$. 
By solving this numerically, we obtain $\phi_k$ and $\rho_k$.   
We have confirmed that $\phi(y)$ within the 2DES  
has little dependence on $W_{\rm p}$ 
if the gap between 2DES strips is wide enough. 
In the following we present results for $W_{\rm p}=2W$.

\subsection{Calculated Results}

We use the following dimensionless variable:  
\begin{equation}
\tilde y = y/l_{\rm U} , 
\end{equation}
with a unit
\begin{equation}
l_{\rm U} = 2 \cyy^{0\rm b} / \ve 
= \nu_{\rm b} l^2 /(\pi a_{\rm B}^*) ,
\label{eq:lp}
\end{equation}
where an estimate using eq.(\ref{eq:estimate}) is substituted for $\cyy^{0\rm b}$. 
When we use $B=5$T, $\nu_{\rm b}=4$ and the value of $m$ and $\ve$ of GaAs, 
we have $l_{\rm U} \sim l$, 
while $l_{\rm U}$($\propto B^{-2}$) becomes larger at smaller $B$. 
We introduce the normalized Hall field and potential as 
\begin{equation}
\tilde E_y = E_y \syy^{0{\rm b}}/ (E_x \syx^{0{\rm b}}) , \ \ \ 
\tilde \phi = \phi \syy^{0{\rm b}}/ (E_x \syx^{0{\rm b}} l_{\rm U}) .  
\end{equation}
From this definition $\tilde E_y=-1$ within the uniform bulk region in the steady state  
since $j_y= \syx^{0{\rm b}} E_x + \syy^{0{\rm b}} E_y =0$. 
From eqs.(\ref{eq:charge_conservation}), (\ref{eq:electrostatic}) 
and (\ref{eq:K_long_range}) we can show that 
$\tilde \phi$ as a function of $\tilde y$ 
is determined only by $\tilde \omega$ and $\tilde W=W/l_{\rm U}$ 
if the edge region is negligible and 
the gap between 2DES strips is wide enough. 

\begin{figure}[ht]
  \begin{center}
  \includegraphics[height=12cm, bb=0 0 595 840]{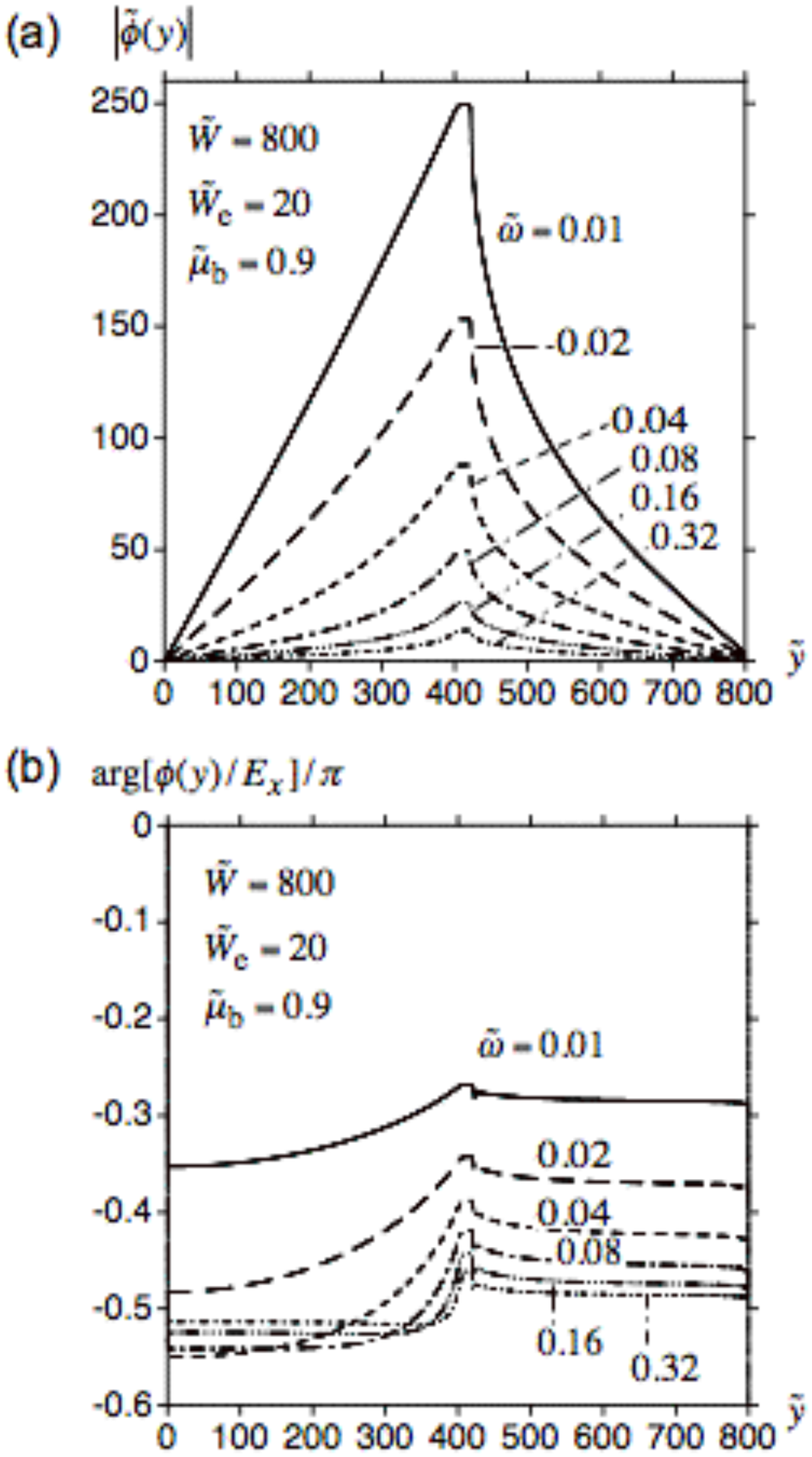}
  \end{center}
  \vskip -0.8cm
  \caption{
  Spatial profile of the Hall potential $\phi(y)$
  in the long-range interaction eq.(\ref{eq:K_long_range}) 
  at several values of $\tilde \omega$ 
  defined by eq.(\ref{eq:omega_tilde}). 
  }
  \label{fig:long_range}
\end{figure}

Figure \ref{fig:long_range}(a) presents 
the absolute value of $\tilde \phi$ in the long-range interaction 
as a function of $\tilde y$ 
for several values of $\tilde \omega$ 
when $\tilde W=800$ and 
the $y$ dependence of $\syx^0$, $\syy^0$ and $\cyy^0$ is given 
as in Fig.\ref{fig:syy}(b).  
Its spatial dependence within the 2DES ($-W/2<y<W/2$) demonstrates 
a crossover from a slope with a constant angle (uniform Hall field) 
to that with a larger angle at both edges compared to the center (concentrated Hall voltage) 
with increasing $\tilde \omega$. 
This crossover in the long-range interaction 
is essentially the same as that obtained in the short-range interaction 
in \S \ref{sec:Short-Range}. 
The argument of $\phi(y)$ shown in Fig.\ref{fig:long_range}(b) exhibits 
a delay relative to that of $E_x$. 
The phase delay is absent at $\tilde \omega=0$, 
increases with increasing $\tilde \omega$, and 
approaches $\pi/2$ at $\tilde \omega \rightarrow \infty$.  
It is shown from eqs.(\ref{eq:charge_conservation}), (\ref{eq:electrostatic}) 
and (\ref{eq:K_long_range}) that, 
as $\tilde \omega \rightarrow \infty$, 
$\phi(y) i\omega /E_x$ approaches a real value independent of $\omega$. 

\begin{figure}[ht]
  \begin{center}
  \includegraphics[height=12cm, bb=0 0 595 840]{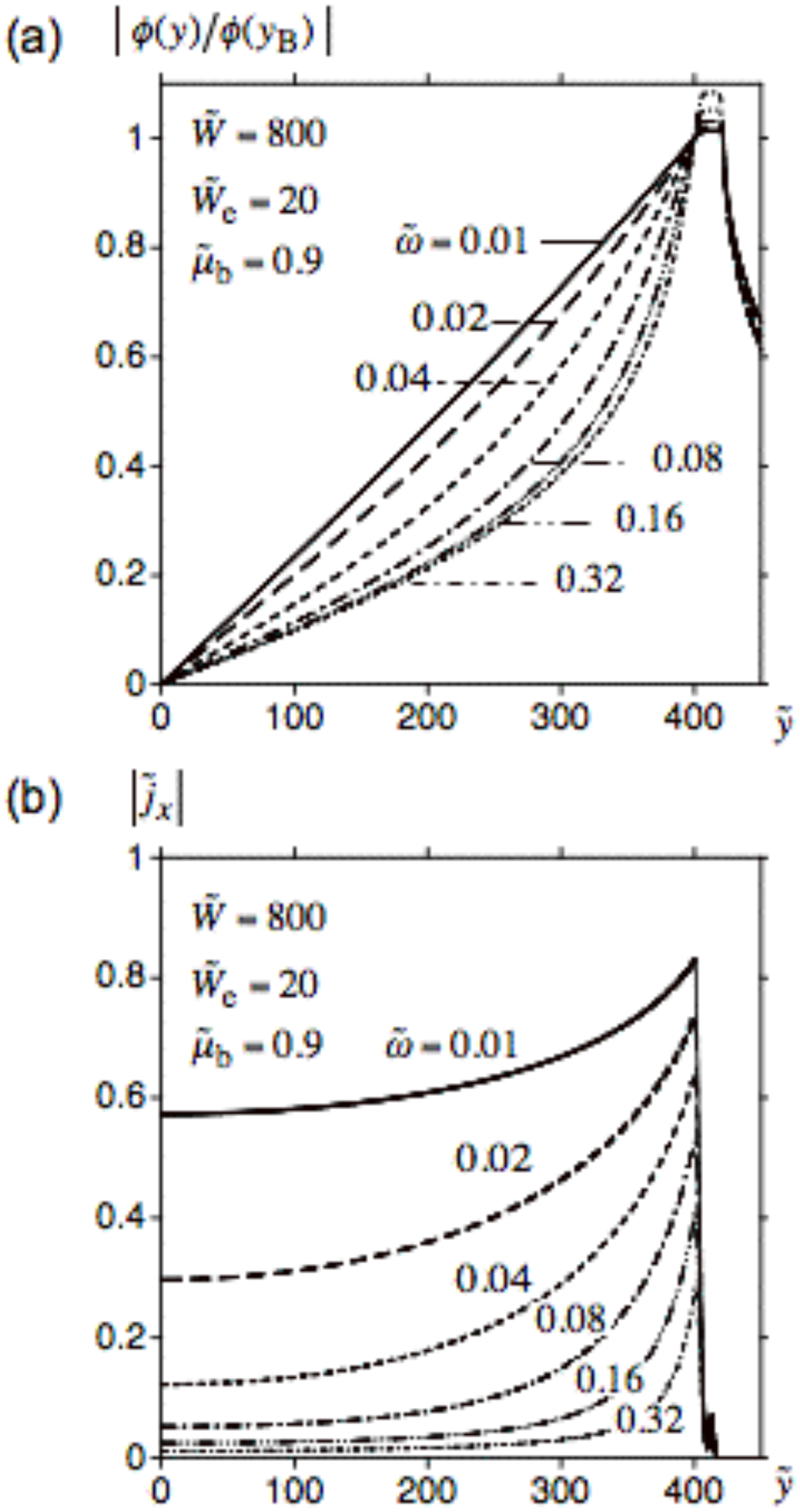}
  \end{center}
  \vskip -0.5cm
  \caption{
    Crossover 
    in the long-range interaction eq.(\ref{eq:K_long_range}),  
    (a) in the Hall potential $\phi(y)/\phi(y_{\rm B})$ 
    with $y_{\rm B}=W/2$ and 
    (b) in the current density $j_x(y)$.
  }
  \label{fig:long_range_at_cross}
\end{figure}

Figure \ref{fig:long_range_at_cross} shows 
calculated results focused on the crossover 
by plotting a normalized Hall potential 
$\phi(y)/\phi(y_{\rm B})$ with $y_{\rm B}=W/2$ 
in Fig.\ref{fig:long_range_at_cross}(a) 
and a normalized current density 
$\tilde j_x(y)=j_x(y) \rho_{xx}^{0{\rm b}}/E_x$ with 
$\rho_{xx}^{0{\rm b}}=\syy^{0{\rm b}}/
(\sxx^{0{\rm b}}\syy^{0{\rm b}}-\sxy^{0{\rm b}}\syx^{0{\rm b}})$
in Fig.\ref{fig:long_range_at_cross}(b). 
The normalized current density $\tilde j_x(y)$ depends on $\syy^{0{\rm b}}/\syx^{0{\rm b}}$. 
In Fig.\ref{fig:long_range_at_cross}(b) 
the value of $\syy^{0{\rm b}}/\syx^{0{\rm b}}=0.01$ is used.  
However, $\tilde j_x(y)$ at such a small value of $\syy^{0{\rm b}}/\syx^{0{\rm b}}$ 
is approximately the same in the bulk uniform region as 
$\tilde j_x(y)$ at $\syy^{0{\rm b}}/\syx^{0{\rm b}}=0$, 
which is equal to $- \tilde E_y$ 
in the bulk uniform region 
where $\syx^0$ takes a constant value $\syx^{0{\rm b}}$. 
Figure \ref{fig:long_range_at_cross}(b) demonstrates 
the crossover in the $y$ dependence of $|j_x|$ and $|E_y|$. 
Although 
$|j_x|$ and $|E_y|$ decrease with increasing $\tilde \omega$ 
also at the boundary $y_{\rm B}=W/2$, 
the decrease is faster in the central region than at the boundary.  
Also note that 
$|j_x|$ and $|E_y|$ at higher $\tilde \omega$ in the long-range interaction 
have a longer tail into the bulk region compared to 
those in the short-range interaction 
which show an exponential decay as derived in \S \ref{sec:Short-Range}. 
Such a longer tail is 
understood from the nonlocal relation between $\phi(y)$ and $\rho(y')$ 
in eq.(\ref{eq:electrostatic}) in the long-range interaction.

\begin{figure}[ht]
  \begin{center}
  \includegraphics[height=12cm, bb=0 0 595 840]{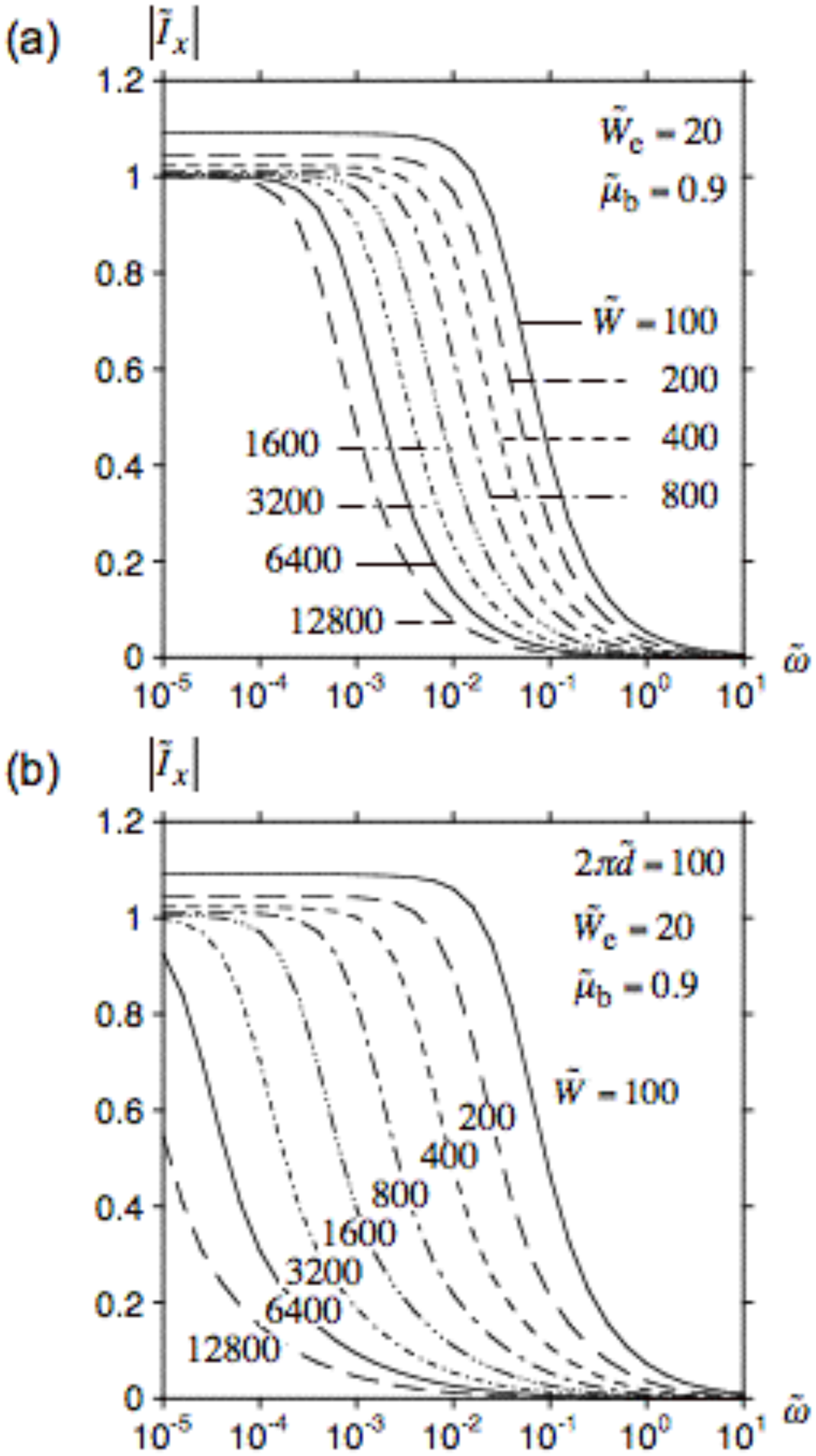}
  \end{center}
  \vskip -0.5cm
  \caption{
  Current $I_x$ as a function of $\tilde \omega$
  (a) in the long-range interaction eq.(\ref{eq:K_long_range})  
  and (b) in the short-range interaction 
  eqs.(\ref{eq:K_short_range})(\ref{eq:rK}).   
  }
  \label{fig:Ix}
\end{figure}

Figure \ref{fig:Ix}(a) presents 
the absolute value of a normalized current  
$\tilde I_x = I_x \rho_{xx}^{0{\rm b}}/(E_x W)$ 
as a function of $\tilde \omega$ 
where $I_x$ is the current per strip defined by
\begin{equation}
I_x = \int_{-W_{\rm p}/2}^{W_{\rm p}/2} dy \ j_x \ .
\end{equation}
Note that $I_x /(E_x W)$ is equal to the inverse of 
\textit{the AC magnetoresistance} $R_{xx}=V_x /I_x$ 
when the distance between the voltage probes is $W$. 
The absolute value of $I_x$ exhibits a drop with increasing $\tilde \omega$.  
The value of $\tilde \omega$ at the drop coincides roughly 
with that at the crossover 
from the uniform to the concentrated distribution, $\tilde \omega_{\rm cross}$,  
while the drop of $I_x$ is not only by 
the reduction of $|E_y(y)|$ in the central region of the strip 
but also by the reduction around $y=\pm y_{\rm B}$. 

Figure \ref{fig:Ix}(a) shows that 
$\tilde \omega_{\rm cross}$ decreases with increasing $W$ as
\begin{equation}
\log \tilde \omega_{\rm cross} \approx -  \log W + {\rm const.} 
\label{eq:omega_cross_long}
\end{equation}
This $W$ dependence of $\tilde \omega_{\rm cross}$ 
in the long-range interaction 
is different from that in the short-range interaction 
shown in Fig.\ref{fig:Ix}(b):  
\begin{equation}
\log \tilde \omega_{\rm cross} \approx - 2 \log W + {\rm const.}
\label{eq:omega_cross_short}
\end{equation}
The analytical expression of $\lambda(\tilde \omega)$ 
in the short-range interaction given by eq.(\ref{eq:decay_length1}) 
leads to $\tilde \omega_{\rm cross} \propto W^{-2}$ 
if we use $\lambda(\tilde \omega_{\rm cross})=W$, 
in agreement with the numerical result eq.(\ref{eq:omega_cross_short}).

\begin{figure}[ht]
  \begin{center}
  \includegraphics[height=12cm, bb=0 0 595 840]{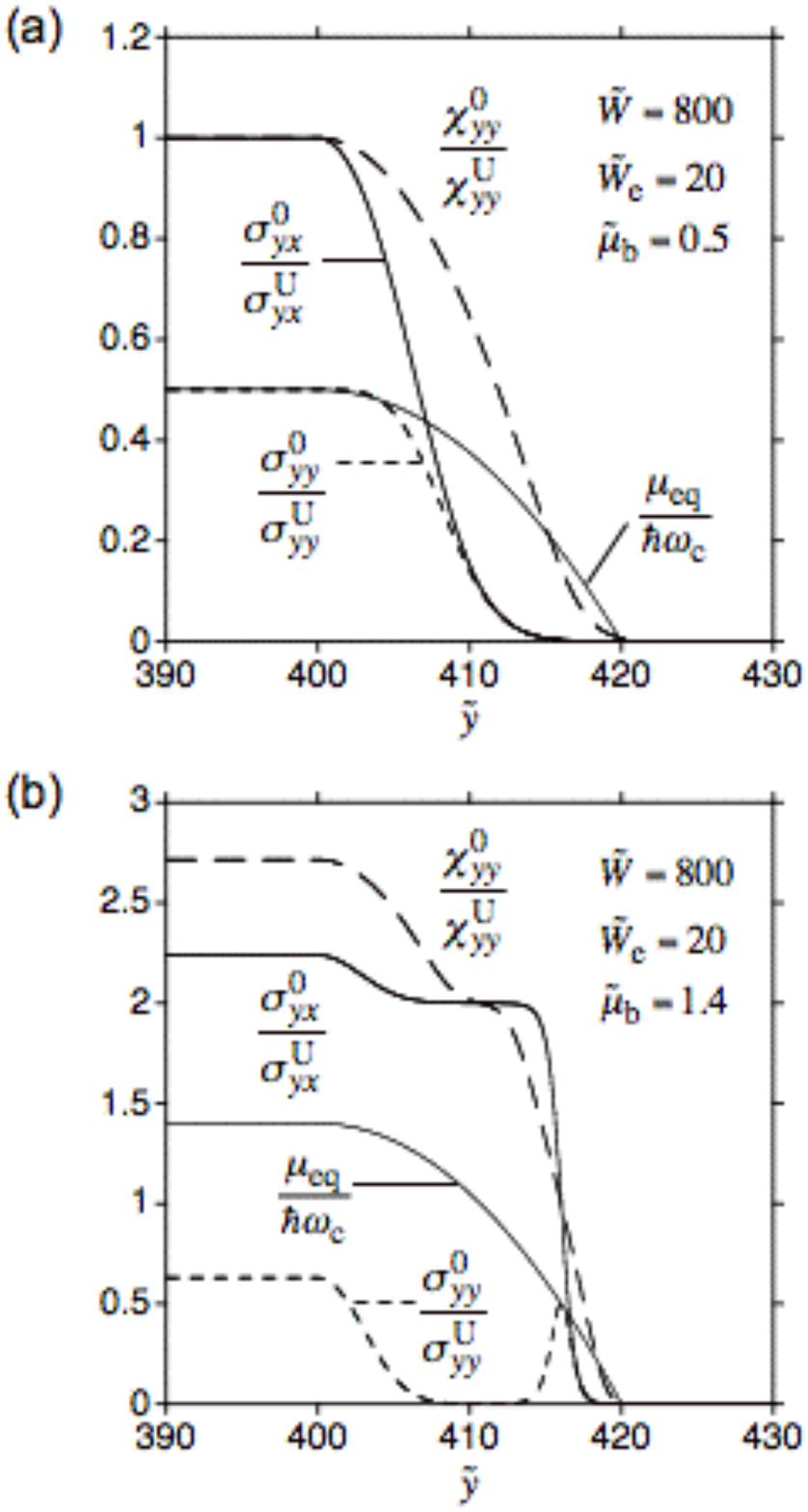}
  \end{center}
  \vskip -0.8cm
  \caption{
  Spatial profiles of $\syx^0$, $\syy^0$ and $\cyy^0$ as well as $\mu_{\rm eq}$  
  for different equilibrium chemical potentials in the bulk region: 
  (a) $\tilde \mu_{\rm b} =0.5$ and 
  (b) $\tilde \mu_{\rm b} =1.4$. 
  }
  \label{fig:syy_mub_dep}
\end{figure}

\begin{figure}[ht]
  \begin{center}
  \includegraphics[height=12cm, bb=0 0 595 840]{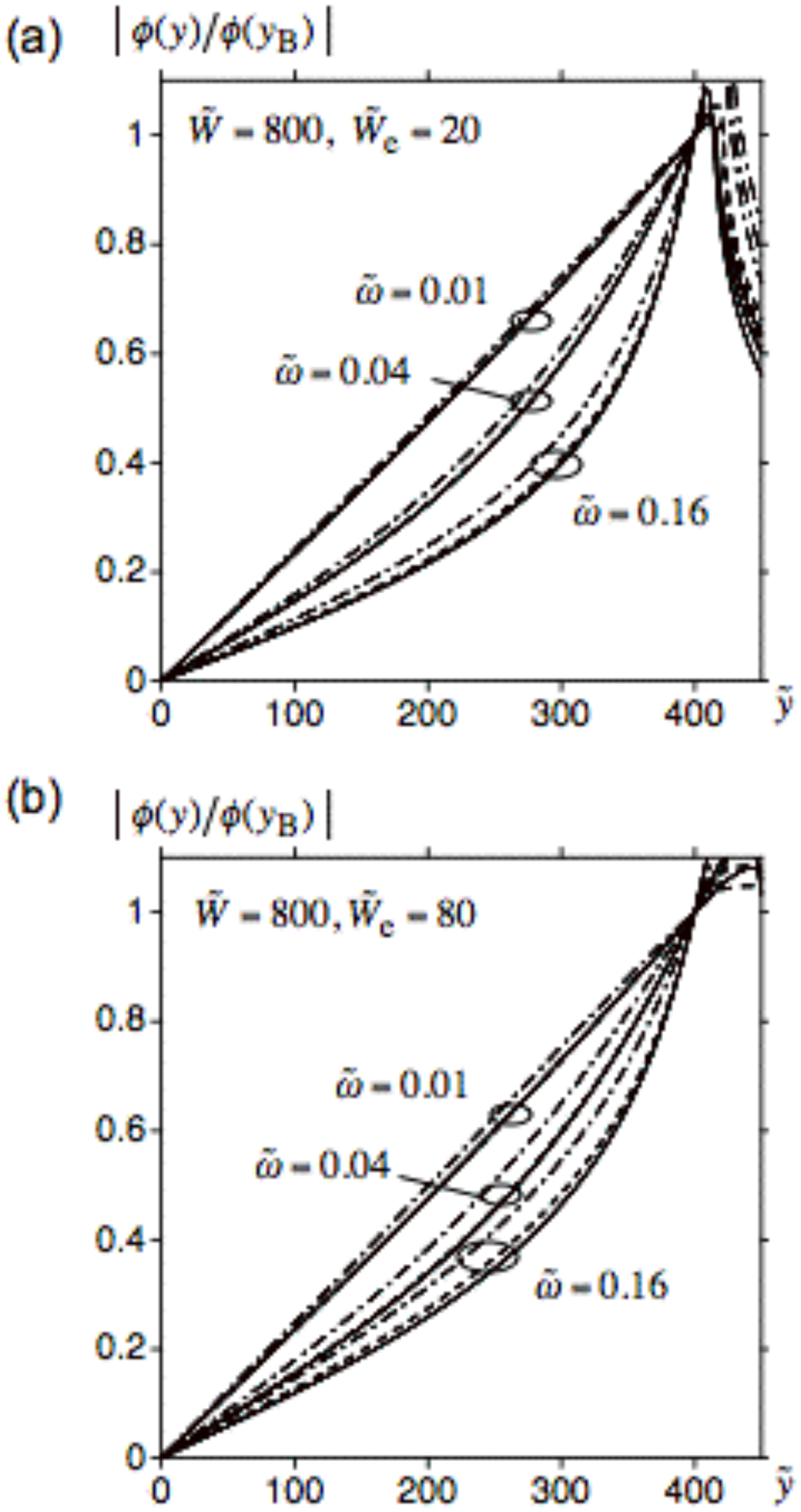}
  \end{center}
  \vskip -0.8cm
  \caption{
  Hall potential $\phi(y)$ 
  in the long-range interaction eq.(\ref{eq:K_long_range})  
  for different equilibrium chemical potentials in the bulk region, 
  $\tilde \mu_{\rm b} =0.5$ ( --------- ), 
  $\tilde \mu_{\rm b} =0.9$ ( $-\ -\ -$ ) and 
  $\tilde \mu_{\rm b} =1.4$ ( $-\cdot -\cdot -\cdot$ ). 
  (a) $\tilde W_{\rm e}=20$, 
  (b) $\tilde W_{\rm e}=80$. 
  }
  \label{fig:long_range_mub_dep}
\end{figure}

Finally we show that 
the crossover from the uniform to the concentrated distribution in the bulk uniform region 
does not change, at least qualitatively, 
when the values of $\syx^0(y)$, $\syy^0(y)$ and $\cyy^0(y)$ in the edge region 
are changed, 
even in the long-range interaction.   
We introduce a variation of $\syx^0(y)$, $\syy^0(y)$ and $\cyy^0(y)$ 
by changing the equilibrium chemical potential in the bulk region as   
$\tilde \mu_{\rm b} =0.5$ (Fig.\ref{fig:syy_mub_dep}(a)), 
$\tilde \mu_{\rm b} =0.9$ (Fig.\ref{fig:syy}(b)) and 
$\tilde \mu_{\rm b} =1.4$ (Fig.\ref{fig:syy_mub_dep}(b)). 
Such a change in $\tilde \mu_{\rm b}$ gives 
a large change in the normalized coefficients $\syx^0(y)/\syx^{0\rm b}$, $\syy^0(y)/\syy^{0\rm b}$ 
and $\cyy^0(y)/\cyy^{0\rm b}$ in the edge region. 
Figure \ref{fig:long_range_mub_dep} shows that the large differences in  
the normalized coefficients in the edge region 
give only small differences in $|\phi(y)/\phi(y_{\rm B})|$ in the bulk region. 
Note that $\phi(y)/\phi(y_{\rm B})$ depends only on 
$\tilde y$, $\tilde \omega$, 
and the normalized coefficients. 
In addition, Fig.\ref{fig:long_range_mub_dep} shows that, 
such differences in $|\phi(y)/\phi(y_{\rm B})|$ 
decrease with decreasing the width of the edge region $\tilde W_{\rm e}$. 

In this paper we have chosen a quite simple model for the $\mu_{\rm eq}$ dependence of $\syx^0$, $\syy^0$ and $\cyy^0$.  
However, differences in $\syx^0(\mu_{\rm eq})$, $\syy^0(\mu_{\rm eq})$ and $\cyy^0(\mu_{\rm eq})$ 
between this model and the more accurate model affect little the crossover, 
since 
we have shown above that $|\phi(y)/\phi(y_{\rm B})|$ in the bulk region is quite insensitive to 
$\syx^0(y)$, $\syy^0(y)$ and $\cyy^0(y)$ in the edge region.

\section{Conclusions and Discussion}

We have studied the Hall potential $\phi(y)$ and  
the Hall field $E_y(y)$ as a function of $y$ (in the width direction)
in quantum Hall systems with width $W$ 
in the case of low-$\omega$ AC current 
in the incoherent linear transport.   
The dynamics of the local Hall-charge density in the uniform bulk region 
is determined by 
the complex diagonal conductivity $\syy= \syy^{0\rm b} + i\omega \cyy^{0\rm b}$ 
where $\syy^{0\rm b}$ and $\cyy^{0\rm b}$ are 
the DC conductivity and the DC dielectric susceptibility, respectively, 
in the bulk. 
We have made calculations in the long-range interaction 
as well as in the short-range interaction, 
and have obtained the following conclusions common to both interactions.  
In the lower-$\omega$ region of 
$\tilde \omega=\omega \cyy^{0\rm b} / \syy^{0\rm b} \ll 1$ 
the transport component is dominant and 
the decay length $\lambda$ of $E_y(y)$ decreases 
with increasing $\omega$. 
When the decay length becomes comparable to $W$ ($\omega = \omega_{\rm cross}$), 
$E_y(y)$ makes a crossover 
from \textit{uniform} to \textit{concentrated-near-edges} profile \cite{Cabo1994}. 
In the higher-$\omega$ region of $\tilde \omega \gg 1$, on the other hand, 
the polarization component is dominant 
and the decay length approaches a constant value.  
The crossover of the Hall-voltage distribution 
at $\omega = \omega_{\rm cross}$ 
is reflected in the frequency dependence of 
the magnetoresistance $R_{xx}(\omega)$.  
With increasing $\omega$ around $\omega_{\rm cross}$,  
$|R_{xx}|$ rises and 
the delay appears in the phase of the current relative to the voltage.  
Note that such a crossover also occurs when 
$\syy^{0\rm b}$ is decreased at a fixed $\omega$. 


The Hall-voltage distribution 
depends on $\omega$ and $\syy^{0\rm b}$ mainly through 
$\tilde \omega=\omega \cyy^{0\rm b} / \syy^{0\rm b}$. 
In the vicinity of $(\omega, \syy^{0\rm b})=(0,0)$ 
we therefore obtain different distributions  
depending on the order of taking 
the limit of $\omega \rightarrow 0$ and that of $\syy^{0\rm b} \rightarrow 0$.  
The decay length $\lambda$ approaches a constant value 
when the limit of $\syy^{0\rm b} \rightarrow 0$ is taken first 
($\tilde \omega \rightarrow \infty$), 
while $\lambda$ becomes infinity 
when that of $\omega \rightarrow 0$ is taken first ($\tilde \omega \rightarrow 0$).  
The theory in the ideal 2DES by MacDonald et al. \cite{MacDonald1983} 
corresponds to the case of $\tilde \omega \rightarrow \infty$ 
since $\syy=i\omega \cyy^0$ in their theory.   
In fact this theory obtains 
the concentrated-near-edges distribution, 
which we have reproduced in the case of $\tilde \omega \rightarrow \infty$. 
On the other hand, the dissipative DC transport giving the uniform distribution
corresponds to the case of $\tilde \omega \rightarrow 0$, 
in which we have reproduced the uniform Hall field.  

Low-frequency admittance has been theoretically studied 
in the edge-channel picture of quantum Hall conductors \cite{Christen1996}, 
in which 
the electrochemical potential of an edge channel 
is equal to that of a contact connected to the channel. 
Since the current through the channel is determined by the distant contact, 
the transport is nonlocal. 
On the other hand,  
this paper is based on the incoherent bulk picture 
in which the electrochemical potential of an edge state 
is considered to be equal to that of the neighboring bulk region. 
In this picture the transport is assumed to be local as in eq.(\ref{eq:ACj}).  

In the experiment by Fontein et al. \cite{Fontein1992} 
the Hall-voltage distribution has been measured 
at a fixed frequency of $\omega/2\pi = 235$Hz 
for two sets of temperature and current values: 
(A) $T=1.5$K, $I=5\mu$A and (B) $T=55$K, $I=20\mu$A. 
In (A) the Hall voltage is concentrated near edges, 
while in (B) it is uniformly distributed. 
The value of $\syy^{0\rm b}$ is much larger in (B). 
If we apply the present theory to interpret this experiment, 
$\tilde \omega$ is decreased with the increase of $\syy^{0\rm b}$ 
and therefore the crossover has occurred from concentrated-near-edges to uniform distribution. 
In this interpretation, 
from the condition that the experimental value of $\tilde \omega$ 
coincides with its theoretical value, 
we can obtain an estimate of $\syy^{0\rm b}$ at the crossover in the experiment, 
which should be between those in (A) and (B). 


Here we make such estimation of $\syy^{0\rm b}$ at the crossover 
from $\tilde \omega_{\rm cross}=\omega \cyy^{0\rm b} / \syy^{0\rm b}$ 
where we use the theoretical value for $\tilde \omega_{\rm cross}$  
and the experimental value for $\omega/2\pi = 235$Hz. 
The sample width $W=2$mm and $l_{\rm U} \sim l \sim 0.01\mu$m give 
$\tilde W=W/l_{\rm U} \sim 10^5$. 
The largest $\tilde W$ at which the crossover has been demonstrated  
in Fig.\ref{fig:Ix}(a) is $\tilde W=12800$, 
for which we have obtained $\tilde \omega_{\rm cross} \sim 10^{-3}$. 
By extrapolating the relation 
$\tilde \omega_{\rm cross} \propto \tilde W^{-1}$ 
in eq.(\ref{eq:omega_cross_long}), 
we have $\tilde \omega_{\rm cross} \sim 10^{-4}$ at $\tilde W=10^5$. 
We use $\cyy^{0\rm b} = e^2 \nu_{\rm b} / (h \oc)$ from eq.(\ref{eq:estimate}), 
$\nu_{\rm b}=4$, $B=5$T and the effective mass of GaAs.  
Then we obtain an estimate of $\syy^{0\rm b} = 10^{-10}\Omega^{-1}$. 
It may not be unrealistic that this value of $\syy^{0\rm b}$ is between the values of $\syy^{0\rm b}$
in (A) and (B). 

Time scales longer than 0.01s 
have been observed in various experiments
in quantum Hall sytems \cite{Weis1998,Kalugin2003,Buss2005,Kershaw2007} 
and 
some of them have already been attributed to small values of $\syy^{0\rm b}$ 
at the time of publication \cite{Weis1998,Kershaw2007}. 
A theory based on small values of $\syy^{0\rm b}$ 
has also been proposed \cite{Akera2009} to explain 
the experiments in the vicinity of the breakdown of 
the quantum Hall effect \cite{Kalugin2003,Buss2005}.

For a contactless 2DES, 
the response to the AC electric field has been studied
with use of capacitively-coupled electrodes in strong magnetic fields 
and a sharp drop of the response with increasing frequency 
has been observed in the MHz region \cite{Grodnensky1991}. 
To explain this drop, 
a theory for a contactless 2DES has been developed which assumes 
the short-range interaction as in eq.(\ref{eq:K_short_range})
and takes into account only the transport component \cite{Grodnensky1991,Grodnensky1992}. 
This theory has derived the length scale of charge accumulation $l_E$, 
which is essentially the same as eq.(\ref{eq:decay_length1}). 
By comparing with the theory, 
the observed drop has been attributed to 
a crossover from bulk to edge response 
which occurs when $l_E$ becomes smaller than the sample size.  

Finally we note that 
the decay length of the Hall electric field (eq.(\ref{eq:decay_length1}))  
and the length scale of charge accumulation \cite{Grodnensky1991,Grodnensky1992}
are the penetration depth in the AC diffusion problem.  
The same penetration depth is encountered 
in the velocity distribution in fluid dynamics (the Stokes layer) \cite{Stokes1851}
and in the temperature distribution in the AC calorimetry \cite{Jeong1997}. 

\section*{Acknowledgment}
The author would like to thank 
T. Ando, H. Suzuura, and A.H. MacDonald for valuable discussions.

\appendix
\section{}

In this Appendix we estimate the value of 
$\cyy^0$ and $\cyx^0$ in eq.(\ref{eq:complex_sigma}) 
by neglecting the Landau-level mixings 
and by neglecting the Landau-level broadening compared to $\hbar \oc$. 

The conductivity $\sigma_{\alpha \beta}$,  
which corresponds to the uniform current density in the $\alpha$ direction 
induced by a uniform electric field with angular frequency $\omega$ 
applied along $\beta$ ($\alpha,\beta=x,y$), 
is expressed by the Kubo formula \cite{Kubo1957,Kubo1959}: 
\begin{equation}
\sab(\omega)=\frac{1}{S} \int_0^{\infty} dt \ e^{-i \omega t - \ve t} 
\int_0^{\beta} d \lambda 
\left< \hat j_{\beta}(-i\hbar \lambda ) \hat j_{\alpha}(t)\right>  ,
\label{eq:Kubo_formula}
\end{equation}
where $S$ is the area of the 2DES, $\ve$ the positive infinitesimal, 
and $\beta = (k_{\rm B} T)^{-1}$ 
with $k_{\rm B}$ the Boltzmann constant and $T$ the temperature. 
The current operator $\hat j_{\alpha} (t)$ in the above equation 
is given by 
$
\hat j_{\alpha} (t) 
=
e^{i \hat H t/\hbar } \hat j_{\alpha}  e^{-i \hat H t/\hbar }
$
,
where $\hat H$ is the Hamiltonian and $\hat j_{\alpha}$ is given by 
\begin{equation}
\hat j_{\alpha} 
= \sum_{\sigma} \int 
\psi_{\sigma}^{\dagger} (\Vec r) (-e v_{\alpha} ) \psi_{\sigma} (\Vec r) d \Vec r,
\end{equation}
where $v_{\alpha}$ is the velocity operator and 
$\psi_{\sigma} (\Vec r)$ is the quantized wave function for spin $\sigma$ 
($\sigma= \uparrow,\downarrow$). 
The bracket $\left< \cdots \right>$ in eq.(\ref{eq:Kubo_formula}) means 
that, for an operator $\hat A$,  
$
\left< \hat A \right> = {\rm tr} (\hat \rho_{\rm eq} \hat A) 
$
with the equilibrium density matrix  
$
\hat \rho_{\rm eq} = e^{-\beta \hat H} /{\rm tr} (e^{-\beta \hat H}) 
$.

We employ the one-electron approximation in which 
\begin{equation}
\hat H = \sum_{\sigma} \int 
\psi_{\sigma}^{\dagger} (\Vec r) H_{\sigma} \psi_{\sigma} (\Vec r) d \Vec r.
\end{equation}
The one-electron operator $H_{\sigma}$ has 
the eigenfunction $\varphi_p(\Vec r)$ 
with $p$ a set of quantum numbers 
and the eigenvalue $\ve_{p\sigma}$  
which satisfy 
$H_{\sigma} \varphi_p(\Vec r) = \ve_{p\sigma} \varphi_p(\Vec r)$. 
We expand $\psi_{\sigma} (\Vec r)$ in terms of the eigenfunctions $\varphi_p(\Vec r)$: 
\begin{equation}
\psi_{\sigma} (\Vec r) = \sum_p c_{p \sigma} \varphi_p (\Vec r),
\end{equation}
and then obtain 
\begin{equation}
\hat H = \sum_{p\sigma} \ve_{p\sigma} c^{\dagger}_{p\sigma} c_{p\sigma}.
\end{equation}
The current operator $\hat j_{\alpha}$ is also expressed as 
\begin{equation}
\hat j_{\alpha} = \sum_{pp'\sigma}  j_{\alpha}^{p'p} 
c^{\dagger}_{p'\sigma} c_{p\sigma},
\end{equation}
with 
\begin{equation}
 j_{\alpha}^{p'p} = \int \varphi^*_{p'}(\Vec r) (-e v_{\alpha}) \varphi_{p}(\Vec r) d \Vec r .
\end{equation}
In such one-electron approximation, we obtain
\begin{equation}
\sab(\omega)=\frac{1}{S} \sum_{p p' \sigma} j_{\beta}^{pp'} j_{\alpha}^{p'p}
T_{p' p \sigma} 
g( E_{p' p \sigma}- \hbar \omega ) ,
\label{eq:sab_omega}
\end{equation}
with
\begin{equation}
\begin{split}
T_{p' p \sigma} 
&= \hbar (f_{p\sigma}-f_{p'\sigma})/E_{p' p \sigma} 
\ \ (p\not=p'), \\
&= \hbar \beta f_{p\sigma} (1- f_{p\sigma}) 
\ \ (p=p'),
\end{split}
\end{equation}
where $E_{p' p \sigma}=\ve_{p'\sigma}-\ve_{p\sigma}$ 
and $f_{p\sigma} = 1/\{\exp[\beta(\ve_{p\sigma}-\mu_{\rm eq})]+1\}$ 
with $\mu_{\rm eq}$ the equilibrium chemical potential, and 
\begin{equation}
g( E )=
i \frac{\mathcal{P}}{E} + \pi \delta (E) .
\end{equation}

First we consider the ideal 2DES where the random potential $V_{\rm ran}=0$. 
In the ideal 2DES the eigenfunction is labeled by $N$ and $k$
where $N=0,1,\cdots$ is the Landau index and $k$ is the momentum along $x$. 
In this case, $j_{\alpha}^{N'k',Nk}$ is diagonal in $k$ and 
\begin{equation}
j_{\alpha}^{N'k,Nk}=0 ,\ \  {\rm except}\  N'=N\pm 1  ,
\end{equation}
and we obtain 
\begin{equation}
\begin{split}
\syy^0&=0  , \ \ \ 
\cyy^0=e^2 \nu/(h \oc)   , \\
\syx^0&=e^2 \nu/h   , \ \ \ 
\cyx^0=0  .
\end{split}
\label{eq:estimate}
\end{equation}

Next we consider the 2DES where $V_{\rm ran} \not= 0$.  
We neglect the Landau-level mixings induced by $V_{\rm ran}$ for simplicity. 
Then the eigenfunction is written as 
\begin{equation}
\phi_{N \gamma} (\Vec r) = \sum_{k} a_{\gamma k}^N \varphi_{Nk} (\Vec r),
\end{equation}
which leads to
\begin{equation}
j_{\alpha}^{N'\gamma',N\gamma}=0 ,\ \  {\rm except}\  N'=N\pm 1  .
\end{equation}
In addition we assume that $\Gamma \ll \hbar \oc$  
where $\Gamma$ is the Landau-level broadening due to $V_{\rm ran}$.   
Then we obtain the same formulas for $\syy^0$, $\cyy^0$, $\syx^0$ and $\cyx^0$  
as those in the absence of $V_{\rm ran}$, eq.(\ref{eq:estimate}), 
except that $\nu$ is the spatial average of the filling factor 
in the presence of $V_{\rm ran}$.
When Landau-level mixings are taken into account, 
$\syy^0$ and $\cyx^0$ become nonzero.  




\begin{thebibliography}{00}

\bibitem{Klitzing1980} 
K. von Klitzing, G. Dorda and M. Pepper: 
Phys. Rev. Lett. \textbf{45} (1980) 494. 

\bibitem{Kawaji1981} 
S. Kawaji and J. Wakabayashi: 
in \textit{Physics in High Magnetic Fields}, edited by S. Chikazumi and N. Miura  
(Springer, Berlin, 1981) p.\ 284.

\bibitem{MacDonald1983} A.H. MacDonald, T.M. Rice and W.F. Brinkman:  
Phys. Rev. B {\bf 28} (1983) 3648.

\bibitem{Thouless1985} 
See, for example, 
D.J. Thouless: J. Phys. C {\bf 18} (1985) 6211.

\bibitem{GrootMazur1962} 
See, for example, 
S.R. de Groot and P. Mazur: 
{\it Non-equilibrium Thermodynamics} 
(North-Holland, Amsterdam, 1962).

\bibitem{Heinonen1985} O. Heinonen and P.L. Taylor:
Phys. Rev. B \textbf{32} (1985) 633. 

\bibitem{Pfannkuche1992} D. Pfannkuche and J. Hajdu:
Phys. Rev. B \textbf{46} (1992) 7032. 

\bibitem{Wexler1994} C. Wexler and D.J. Thouless:
Phys. Rev. B \textbf{49} (1994) 4815. 

\bibitem{Thouless1993} D.J. Thouless:
Phys. Rev. Lett. \textbf{71} (1993) 1879. 

\bibitem{Hirai1994} H. Hirai and S. Komiyama:
Phys. Rev. B \textbf{49} (1994) 14012. 

\bibitem{Palacios1998} J.J. Palacios and A.H. MacDonald:
Phys. Rev. B \textbf{57} (1998) 7119. 

\bibitem{Guven2003}
K. G\"uven and R. R. Gerhardts: 
Phys. Rev. B {\bf 67} (2003) 115327.

\bibitem{Siddiki2004}
A. Siddiki and R. R. Gerhardts: 
Phys. Rev. B {\bf 70} (2004) 195335.

\bibitem{Kanamaru2006} S. Kanamaru, H. Suzuura and H. Akera:  
J. Phys. Soc. Jpn. \textbf{75} (2006) 064701.

\bibitem{Fontein1991}
P.F. Fontein, J.A. Kleinen, P. Hendriks, F.A.P. Blom, J.H. Wolter, 
H.G.M. Lochs, F.A.J.M. Driessen, L.J. Giling and C.W.J. Beenakker: 
Phys. Rev. B \textbf{43} (1991) 12090.

\bibitem{Fontein1992}
P.F. Fontein, P. Hendriks, F.A.P. Blom, J.H. Wolter, L.J. Giling and C.W.J. Beenakker: 
Surf. Sci.  \textbf{263} (1992) 91.

\bibitem{Ando1996} T. Ando: 
Surf. Sci.  \textbf{361-362} (1996) 270.  
\bibitem{Ando1998} T. Ando: 
Physica B \textbf{249-251} (1998) 84.  

\bibitem{Akera2005}
H. Akera and H. Suzuura: J. Phys. Soc. Jpn. \textbf{74} (2005) 997.

\bibitem{Cabo1994} 
It has been pointed out without calculation that such a crossover to the uniform distribution occurs when the time interval, in which the current is applied, becomes equal to the relaxation time, by 
A. Cabo and A. Gonz{\' a}lez: 
Revista Mexicana de F{\'i}sica \textbf{40} (1994) 71. 

\bibitem{Christen1996}
T. Christen and M. B{\"u}ttiker: 
Phys. Rev. B \textbf{53} (1996) 2064.  

\bibitem{Weis1998}
J. Weis, Y.Y. Wei and K. v. Klitzing: Physica B \textbf{256-258} (1998) 1. 

\bibitem{Kalugin2003}
N. G. Kalugin, B. E. Sa$\check{\rm g}$ol, A. Buss, A. Hirsch, C. Stellmach, G. Hein and G. Nachtwei: Phys. Rev. B \textbf{68} (2003) 125313.

\bibitem{Buss2005}
A. Buss, F. Hohls, F. Schulze-Wischeler, C. Stellmach, G. Hein, R. J. Haug and G. Nachtwei: Phys. Rev. B \textbf{71} (2005) 195319.

\bibitem{Kershaw2007}
T.J. Kershaw, A. Usher, A.S. Sachrajda, J. Gupta,
Z.R. Wasilewski, M. Elliott, D.A. Ritchie and M.Y. Simmons: 
New Journal of Physics \textbf{9} (2007) 71. 

\bibitem{Akera2009}
H. Akera: 
J. Phys. Soc. Jpn. \textbf{78} (2009) 023708.

\bibitem{Grodnensky1991}
I.M. Grodnensky, D. Heitmann, K. von Klitzing and A.Y. Kamaev: 
Phys. Rev. B \textbf{44} (1991) 1946.

\bibitem{Grodnensky1992}
I.M. Grodnensky, D. Heitmann, K. von Klitzing and A.Y. Kamaev: 
in \textit{High Magnetic Fields in Semiconductor Physics}, edited by G. Landwehr 
(Springer, Berlin, 1992) p.\ 135.

\bibitem{Stokes1851}
G.G. Stokes: Trans. Camb. Phil. Soc. \textbf{9} (1851) 8.

\bibitem{Jeong1997} 
For example, Y.H. Jeong: Thermochimica Acta \textbf{304/305} (1997) 67.

\bibitem{Kubo1957}
R. Kubo: J. Phys. Soc. Jpn. \textbf{12} (1957) 570.
\bibitem{Kubo1959}
R. Kubo, H. Hasegawa, and N. Hashitsume: 
J. Phys. Soc. Jpn. \textbf{14} (1959) 56.

\end{thebibliography}
\end{document}